\documentclass[prd,aps,twocolumn,nofootinbib,floatfix]{revtex4}
\usepackage{color}
\usepackage{graphicx}
\usepackage{amsmath}

\def\gr{$\gamma$-ray}
\usepackage{hyperref}
\begin{document}
\newcommand{\alberto}[1]{\textcolor{blue}{#1}}
\newcommand{\cc}[1]{\textcolor{magenta}{chiara: #1}}
\newcommand{\OmGW}{\Omega_{\rm GW}}
\interfootnotelinepenalty=10000

\title{NANOGrav signal from magnetohydrodynamic turbulence at the QCD phase transition in the early Universe}
\author{Andrii Neronov$^{1,2}$, Alberto Roper Pol$^{1,3}$,  Chiara Caprini$^{1}$ and Dmitri Semikoz$^1$ }
\affiliation{$^1$Université de Paris, CNRS, Astroparticule et Cosmologie,  F-75006 Paris, France\\
$^2$Astronomy Department, University of Geneva, Ch. d'Ecogia 16, 1290, Versoix, Switzerland\\
$^3$ Faculty of Natural Sciences and Medicine, Ilia State University, 0194 Tbilisi, Georgia}

\begin{abstract}
The NANOGrav collaboration has recently reported evidence for the existence of a stochastic gravitational wave background in the 1-100 nHz frequency range. We argue that such background could have been produced by magneto-hydrodynamic (MHD) turbulence at the QCD scale. From the NANOGrav measurement one can infer the magnetic field parameters: comoving field strength close to microGauss and a correlation length close to 10\% of the Hubble radius at the QCD phase transition epoch. We point out that the turbulent decay of a  non-helical magnetic field with such parameters leads to a magnetic field at the recombination epoch, which would be sufficiently strong to provide a solution to the Hubble tension problem, as recently proposed. We also show that the MHD turbulence interpretation of the NANOGrav signal can be tested via measurements of the relic magnetic field in the voids of the large scale structure, with gamma-ray telescopes like CTA.
\end{abstract}
\maketitle

\section{Introduction}
The problem of the origin of cosmic magnetic fields is one of the long-standing problems of astrophysics and cosmology (see, e.g.,~\cite{kronberg}). Magnetic fields in galaxies and galaxy clusters are produced via dynamo amplification of pre-existing, weak, seed magnetic fields, which should have been produced before galaxy formation, possibly in the early Universe. These seed fields may still be found in their original form in the intergalactic medium \cite{kronberg,durrer13,subramanian}. 

Several techniques, such as the observation of Faraday rotation in distant radio sources (see, e.g.,~\cite{kronberg}), the analysis of anisotropies and polarization of the Cosmic Microwave Background (CMB) (see, e.g.,~\cite{planck,planck_h0}), the search for the emission from electromagnetic cascades developing along the line of sight toward distant  \gr\ sources  (see, e.g.,~\cite{neronov09,neronov10,fermi}), have been previously  used to constrain the strength and correlation length of cosmological magnetic fields and to provide information on their origin. 

New opportunities to probe cosmological magnetic fields are emerging, thanks to the development of the new generation of gravitational wave (GW) detectors. The magnetic fields that might have existed in the early Universe are expected to have turbulent structure, and to produce space-time varying tensor stresses, leading to a GW signal \cite{Durrer:1999bk,Caprini:2001nb}. The GWs produced by magnetic fields generated at cosmological phase transitions, are expected to have frequencies in the nHz to mHz range accessible to LISA \cite{Audley:2017drz} and  pulsar timing arrays (PTA) \cite{nanograv11yr,epta}.  

\begin{figure}
\includegraphics[width=\linewidth]{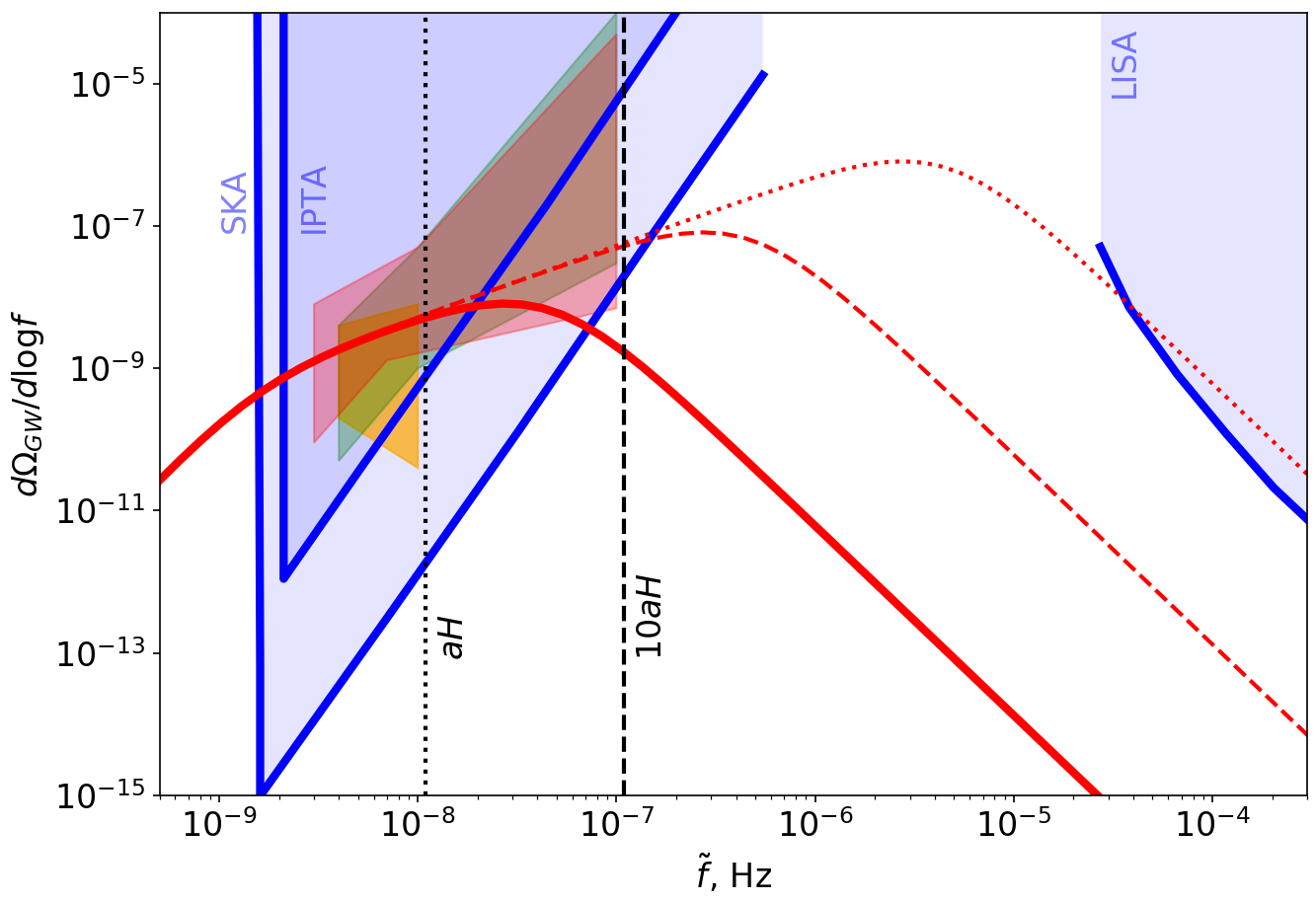}
\caption{Spectra of the GW energy density for GW signal from magnetic fields generated at the QCD phase transition, compared to  IPTA, SKA \cite{ska}, and LISA \cite{lisa_sensitivity} sensitivities. Green and orange wedges correspond to the allowed range of slopes and normalisation of the GW density from the broken power law and power-law fits to the NANOGrav cross-power spectral density \cite{nanograv}. Red wedge is the result of EPTA (3rd Annual Assembly GdR ``Ondes Gravitationnneles,''
\cite{GdR}).
Red solid, dashed and dotted curves show broken power-law type models of the type derived in Ref. \cite{pol19} for different magnetic field forcing scales. 
}

\label{fig:power}
\end{figure}

The NANOGrav Collaboration has reported detection of a signal in the frequency range between 3 and 100 nHz  consistent with a stochastic gravitational wave background (SGWB)\footnote{The NANOGrav observation favors a quadrupolar correlation over monopolar, which indicates that the signal corresponds to GWs \cite{nanograv}.} \cite{nanograv}. 
Figure \ref{fig:power} shows the NANOGrav detection expressed in terms of the frequency spectrum of the density fraction of the SGWB, ${\rm d}\OmGW/{\rm d}\, \mbox{log}f$
[we have used Eq.~(17) from Ref.~\cite{Magg00} to convert the result of Ref.~\cite{nanograv} in this format]. Green and orange wedges in Fig. \ref{fig:power} show the $90\%$ confidence contours of the  $A_{CP},\gamma_{CP}$ parameter space of Ref. \cite{nanograv}, in terms of $(f\,,\,{\rm d}\OmGW/{\rm d}\, \mbox{log}f)$: these have been obtained by fitting a power-law (green) and broken power-law (orange) spectra to the cross-power spectral density of the GW signal.

The conventional models of the SGWB signal are astrophysical: the population of merging super-massive black holes binaries, see, e.g.,~\cite{nanograv,Ding:2020bnl,Xin:2020owo}. Alternative ``new physics'' interpretations of the signal have also been considered, including  cosmic strings  \cite{ellis,Blasi:2020mfx,Buchmuller:2020lbh}, primordial black holes \cite{Vaskonen:2020lbd,DeLuca:2020agl,Bhaumik:2020dor,Kohri:2020qqd} and dark phase transitions \cite{Nakai:2020oit,Addazi:2020zcj}.

In this paper we discuss another ``new physics'' interpretation; i.e., the possibility that the SGWB in the NANOGrav frequency range is due to magnetohydrodynamic (MHD) processes in the early Universe during the epoch of the quantum chromodynamic (QCD) phase transition (see, e.g.,~\cite{hogan,caprini09,qcdpt_tina,Anand:2017kar,Kalaydzhyan:2014wca}). We study the implications of such an interpretation of the NANOGrav signal for the physics of cosmological magnetic fields. We also discuss possible ``multimessenger'' tests of the hypothesis that this SGWB is produced by MHD at the QCD phase transition, using CMB and gamma-ray data. We show that the estimate of the intensity of the magnetic
field produced at the QCD phase transition, which we 
infer from the NANOGrav detection, is consistent with that obtained for the magnetic field at the recombination epoch, based on CMB data \cite{pogosyan}. The postrecombination magnetic field, relic of the QCD epoch, can still reside in the voids of the large scale structure, where it is detectable with gamma-ray telescopes \cite{Korochkin:2020pvg}.
\vskip0.2cm

\section{Estimate of the gravitational wave production by primordial magnetic fields}
The GW energy density power spectrum produced by a primordial magnetic field depends strongly on the space-time structure of the magnetic anisotropic stresses. Therefore, the details of the magnetic field generation mechanism, as well as of the associated MHD turbulence, are expected to play a role in shaping the GW signal. 
Several predictions exist in the literature, that provide different estimations of the SGWB spectrum \cite{Caprini:2001nb,Caprini:2006jb,Gogoberidze:2007an,Kahniashvili:2008pe,Kahniashvili:2009mf,Caprini:2009pr,caprini09,caprini10,sigl18,pol19}. Since the properties of the GW source, such as its evolution and space-time correlation, are determinant for the GW signal, different assumptions on the source dynamics reflect in as many predictions for the SGWB amplitude and shape.  

An estimate of the scaling of the GW energy density with the magnetic field parameters at scales larger than the timescale of the sourcing process can be obtained as follows. GWs are transverse traceless perturbations $h_{ij}$ of the metric  $ds^2 = a^2 \left(d\eta^2 - ( \delta_{ij}+h_{ij} )dx^idx^j \right)$, where $a$ is the scale factor and $\eta$ is the conformal time. Fourier components of $\tilde h_{ij}=ah_{ij}$ satisfy the wave equation (we drop the indices ${ij}$),
\begin{equation}
\label{eq:osc}
\partial^2_\eta \tilde h+ k^2\tilde h=\frac{16\pi G}{a} \tilde{\cal T}^{TT},
\end{equation}
where $G$ is Newton's constant, $k$ is the comoving
wave number, and $\tilde {\cal T}^{TT}=a^4{\cal T}^{TT}$ are the transverse traceless components of the source stress-energy tensor\footnote{We are using the system of units in which the speed of light $c=1$.}. Equation (\ref{eq:osc}) describes the dynamics of a forced oscillator, the source term of which depends on the  magnetic field strength,  $\tilde {\cal T}^{TT}\sim \tilde 
B^2/2 + \tilde \rho (1 + 4 v^2)/3$, with $\tilde B$ and $\tilde\rho$ being the comoving magnetic field
strength and background energy density, and $v$ the plasma velocity.
We suppose that the energy density of the magnetic field is larger than the kinetic energy density of the plasma. $\tilde {\cal T}^{TT}$ is a space-time dependent convolution of the magnetic field; for this order of magnitude estimate, we model it as a constant forcing, operating on a comoving time interval $\tilde\tau$. The solution of equation (\ref{eq:osc}) at the end of the sourcing process, with initial conditions $\tilde h= \partial_\eta \tilde h = 0$, is $[\partial_\eta \tilde h](\eta_*+\tilde\tau) \sim 8 \pi G \tilde B^2 \sin(k\tilde\tau)/ (a_*k) \rightarrow 8 \pi G \tilde B^2 \tilde\tau/a_*$, in the large scale limit $k\tilde\tau<1$ (a star denotes the time at which the GW source starts operating). The comoving energy density of GWs at the end of the sourcing process becomes then $\tilde \rho_{GW}= [\partial_\eta \tilde h]^2/(32\pi G)\simeq 2
\pi G  \tilde B^4 \tilde\tau^2/a_*^2$.
Dividing by the overall comoving density of the Universe $\tilde\rho$ we find the density fraction of GWs: $\OmGW\sim 
3 \,\Omega_B^2 \, (H_*\tau)^2$,
where we have introduced the magnetic energy
density fraction $\Omega_B = B^2/2 \tilde \rho$, and $H_*$ is the Hubble rate at time $\eta_*$. 

The scaling of $\OmGW$, interpreted as the value at the peak of the GW energy density spectrum, is in agreement with that given in Ref.~\cite{Caprini:2019egz}, for the case of a source which has typical duration $\tau$ comparable to its decorrelation time $\tau_c$. 
In the case of GWs generated by MHD turbulence, one expects $\tau\sim \tau_c\sim l_*/v_A\leq (H_*)^{-1}$, where $l_*$ is the flow length-scale at the source time and/or the magnetic field correlation scale, and $v_A=\sqrt{2\Omega_B}$ is the Alfv\'en speed. This leads to 
$\OmGW\sim 
(3/2)\,\Omega_B \, (H_* l_*)^2\,.$
Among the most recent analyses of GW from MHD turbulence, we note that Ref.~\cite{sigl18} predicts a somewhat different scaling $\OmGW\sim \Omega_B^{3/2} \, (H_*l_*)^2$, and the numerical simulations of Ref.~\cite{pol19} find instead
$\OmGW \sim \Omega_B^{2} \, (H_*l_*)^2$. 
A detailed evaluation of the GW signal from MHD turbulence, supported by further numerical simulations and study of the initial conditions, will be the matter of a subsequent work. Here, we simply quantify our ignorance of the correct scaling by introducing a parameter $1\leq n\leq 2$, such that $\OmGW \sim \Omega_B^n (H_* l_*)^2$. As we shall see, this does not affect the result in a relevant manner. Rescaling the GW density fraction to the present day Universe gives
$\Omega_{{\rm GW},0}=(a^4H_*^2/H_0^2)\OmGW$,
where $H_0 =100 h$ km/s/Mpc is the present day expansion rate. We get 
\begin{equation}
\label{eq:omega}
h^2\Omega_{{\rm GW},0}\sim 3.5 \times 10^{-5}\left[\frac{N_{\rm eff}}{10}\right]^{-\frac{1}{3}}\Omega_{B}^n \,(H_*l_*)^2\,,
\end{equation}
for the effective number of relativistic degrees of freedom $N_{\rm eff}\sim 10$ at the temperature $T\sim 100$~MeV.

The characteristic scale of MHD turbulence at a given Hubble time $t_H=H_*^{-1}$ is that of the largest processed eddies, i.e., the eddies for which the turnover time scale is equal to $t_H$. Their length scale is
$
l_{\rm LPE}=v_A/H_*=\sqrt{2\Omega_B}/H_*\,.
$
If we  assume $l_*=l_{\rm LPE}$ in Eq.~\eqref{eq:omega}, we obtain the highest amplitude for the SGWB signal,
\begin{equation}
\label{eq:lpe}
h^2 \Omega_{{\rm GW},0}\sim 7 \times 10^{-5}\left[\frac{N_{\rm eff}}{10}\right]^{-\frac{1}{3}}\Omega_B^{n+1}\,.
\end{equation}

\begin{figure}
\includegraphics[width=\linewidth]{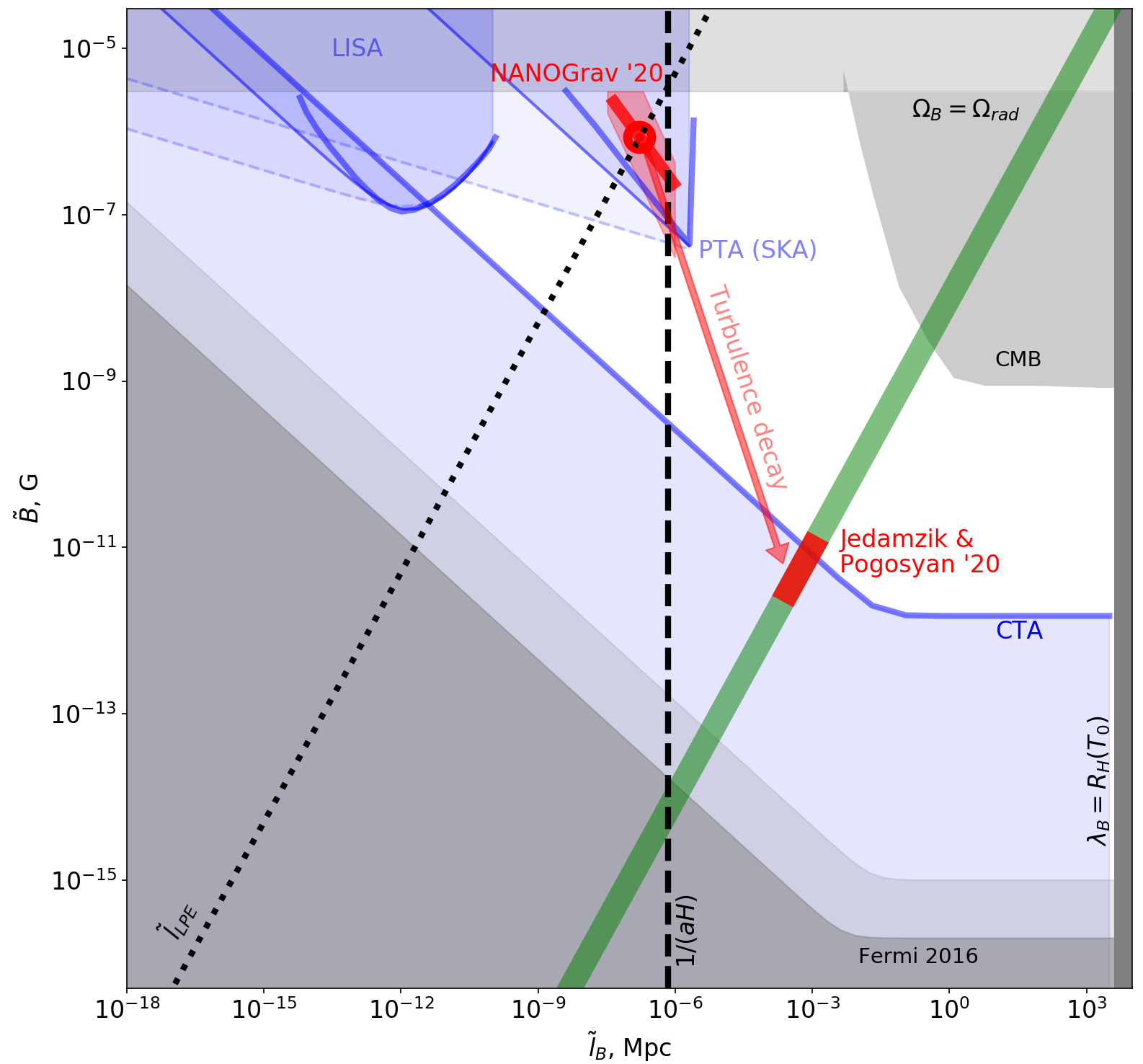}
\caption{Constraints on cosmological magnetic fields, compared to the NANOGrav result \cite{nanograv} interpreted as evidence for the detection of a primordial magnetic field. The sensitivities of GW detectors \cite{nanograv11yr,ska,lisa_sensitivity}  are shown by the upper blue shading for: a sharply peaked SGWB (thick solid), a SGWB with $k^1$ slope (dashed thin), and with $k^3$ slope (thin solid lines). 
The lower light blue shading shows the sensitivity of CTA \cite{neronov09,durrer13,Korochkin:2020pvg}. The lower bounds with different degrees of grey shading show the Fermi/LAT lower bound on the intergalactic magnetic field from timing of the blazar signal (darker), and from the search of extended emission (lighter) \cite{fermi}. The CMB upper bound is from the analysis of Refs. \cite{planck,durrer13}.  The suggested detection of a SGWB is marked by the red circle. Red shading corresponds to the uncertainty range. The green line shows  endpoints of the cosmological evolution of primordial magnetic fields,  at the recombination epoch \cite{banerjee,durrer13}. The red interval on the green line shows the possible detection from Ref.~\cite{pogosyan}. The rose arrow shows the evolutionary path  via compressible turbulent decay \cite{durrer13}. The black dashed and dotted lines show the comoving  Hubble radius and the largest processed eddy at the QCD phase transition.}
\label{fig:exclusion}
\end{figure}

This relation provides a convenient estimate for the  sensitivities of gravitational wave detectors to primordial magnetic fields at the moment of cosmological magnetogenesis. A detector sensitive at the frequency $\tilde f$ can measure the energy fraction of gravitational waves produced by magnetic field modes variable on  the comoving distance scale\footnote{The source term in the wave equation (\ref{eq:osc}) is quadratic in the magnetic field. We consider therefore that the GWs sourced by a magnetic field with correlation length $\tilde l_B$ have wave numbers $k=4\pi/\tilde l_B$.} $\tilde l_B=2/\tilde f$. 
Converting the sensitivity limits of LISA \cite{lisa_sensitivity} and NANOGrav \cite{nanograv11yr}  for ${\rm d}\OmGW/{\rm d} \, \mbox{log}f$ into a limit on $\Omega_B$ using Eq.~(\ref{eq:lpe}) for quadratic scaling with $\Omega_B$, i.e., $n = 2$, we arrive at the sensitivity curves shown in Fig.~\ref{fig:exclusion}.

The quoted sensitivity range of NANOGrav is $2.5\times 10^{-9}\, {\rm Hz}<\tilde f<7\times 10^{-8} \, {\rm Hz}$ \cite{nanograv}, i.e., NANOGrav is sensitive to the GW signal generated by magnetic fields with comoving correlation length in the range $0.1\mbox{ pc}<\tilde l_B< 10$~pc. This range contains the Hubble scale of the QCD phase transition at  $T\sim 100$~MeV
$
    \tilde l_H=(aH)^{-1}\simeq 1\mbox{ pc}\left[N_{\rm eff}/10\right]^{-1/6}\left[T/100\mbox{ MeV}\right]^{-1}\,.
$
\vskip0.2cm

\section{NANOGrav signal from MHD turbulence at the QCD phase transition}
Figure \ref{fig:power} shows the range of $\Omega_{{\rm GW},0}$ values suggested by the NANOGrav measurements  for the range of frequencies covered by the experiment (green and orange shaded regions). We compare this measurement with examples of the GW spectrum $\Omega_{{\rm GW},f}={\rm d}\OmGW/{\rm d}\, \mbox{log} f$ expected from magnetic fields generated at the QCD phase transition. 

As previously mentioned, several estimates of the SGWB signal from MHD turbulence have previously been derived, not always in agreement, following different assumptions in modelling the source properties~\cite{caprini09,Kahniashvili:2009mf,caprini10,sigl18,pol19}. Fairly generally, the SGWB spectrum is expected to be a broken power-law, following a $\tilde f^3$ scaling at small (super-horizon) frequencies $\tilde f < \tilde f_H=aH$ \cite{causality}, possibly breaking to $\tilde f^\alpha$, with $\alpha\simeq 1$ \cite{pol19} or
$\alpha=2$ \cite{caprini10}, or $0<\alpha\leq 3$ \cite{sigl18} in the range of frequencies from $\tilde f \sim \tilde f_H$ to the 
higher frequency scale $f_*\simeq 2/l_*$, which
corresponds to the largest processed eddy scale
or to a smaller forcing scale characteristic of the magnetic field generation process.
In the frequency  range $\tilde f>\tilde f_*$, one expects another power law  $\tilde f^\beta$, with $\beta<0$, which follows from the properties of freely decaying turbulence. The slopes $\alpha$ and $\beta$ depend on the magnetic field initial conditions, on the MHD turbulence type, and on its temporal evolution and decorrelation  \cite{caprini09,sigl18,pol19}. Note that, if the QCD phase transition is first order \cite{QCDpt,firstorderQCD,Wygas:2018otj}, and proceeds through the nucleation of broken phase bubbles, one also expects a GW signal from bubble collisions \cite{Kosowsky:1991ua,Kosowsky:1992vn,Huber:2008hg,Cutting:2018tjt,Cutting:2020nla} and sound waves \cite{Hindmarsh:2013xza,Hindmarsh:2015qta,hindmarsch17,Cutting:2019zws,Hindmarsh:2019phv}. 

Figure \ref{fig:power} shows examples of broken power-law spectra of $\Omega_{{\rm GW},f}$ for different values of the magnetic field forcing scale $f_*$. We have plotted here the SGWB model derived from the numerical simulations of Ref. \cite{pol19}, inserting a causal $f^3$ slope on scales larger than the Hubble scale, not reached by the simulations.  

The NANOGrav measurement suggests the value $h^2\Omega_{{\rm GW},0}\sim 10^{-9}$ at $\tilde f_{\rm year}\sim 3 \times 10^{-8}$~Hz (see
orange wedge of Fig.~\ref{fig:power}), with roughly an order-of-magnitude uncertainty at the $90\%$ confidence level.
Assuming that the spectrum $\Omega_{{\rm GW},f}$ is a power-law with slope $\tilde f^\alpha$ with $0<\alpha\leq 3$ in the frequency range $\tilde f_H<\tilde f<\tilde f_*$, as suggested by analyses of the SGWB from MHD turbulence \cite{caprini10,pol19}, we can derive an order-of-magnitude relation from the NANOGrav measurement,
\begin{equation}
\label{eq:ng}
\frac{h^2\Omega_{{\rm GW},0}/10^{-9}}{[\tilde f/(3\times 10^{-8}\mbox{ Hz})]^\alpha}\simeq 1\,.
\end{equation}
This relation can be used to infer the range of magnetic field strength and correlation length suitable to fulfill the NANOGrav observation. 

The characteristic frequency of the SGWB signal is expected to correspond to the energy-containing scale of the MHD turbulence, 
$
\tilde f\sim 2/l_*=2\times
10^{-8}(H_*l_*)^{-1}\left[N_{\rm eff}/10\right]^{1/6}\left[T/100\mbox{ MeV}\right]\mbox{ Hz}
$.~Substituting this frequency estimate into Eq.~(\ref{eq:ng}), expressing  $\Omega_{{\rm GW},0}$ through $\Omega_B$ using  Eq. (\ref{eq:omega}), and 
setting $n=2$, $\alpha=1$ as indicated by the numerical simulations of Ref.~\cite{pol19},
we find 
\begin{equation}\label{eq:Omfin}
\Omega_B\simeq 
4.5 \times 10^{-3} (H_*l_*)^{-\frac{3}{2}}
\left[\frac{N_{\rm eff}}{10}\right]^{\frac{1}{4}}
\left[\frac{T}{100\mbox{ MeV}}\right]^{\frac{1}{2}}\,.
\end{equation}
This result is shown by the thick red line  in Fig. \ref{fig:exclusion}, and the red shading around the line shows the uncertainty range of the estimates for $0<\alpha\leq 3$ and $1\leq n\leq 2$.  It is clear that the uncertainty in these parameters does not affect in a significant way the value of the magnetic density parameter $\Omega_B$, with respect to observational constraints and to the  range of endpoints of evolution. 

For the special case of the largest processed eddies $l_*=l_{\rm LPE}$, the NANOGrav measurement suggests
$\Omega_B\simeq 0.03$ for $\alpha=1$ and $n=2$, corresponding to the specific values of the magnetic field amplitude $\tilde B$ and correlation scale $\tilde l_B$  
$
    \tilde B\simeq 0.7\, \mu\mbox{G,             }$ 
 $\tilde l_B\simeq 0.2\mbox{ pc}$
shown by the red circle in Fig. \ref{fig:exclusion}. Note that, in order to explain the NANOGrav detection, the energy containing scale of the MHD turbulence, that we set here corresponding to $\tilde l_B$, cannot be too far from the horizon. The magnetic field density parameter must in fact satisfy $\Omega_B\lesssim 0.2$ from the nucleosynthesis bound \cite{Kawasaki:2012va}, leading to $H_*l_* \gtrsim 0.1$ from Eq.~\eqref{eq:Omfin}. 
\vskip0.2cm
\section{Discussion}
The strength and correlation length of a cosmological magnetic field derived from the NANOGrav measurement is well within the range of expectation of the models of magnetic field generation at a first-order QCD phase transition (see, e.g.,~\cite{sigl97,qcdpt_tina}). In these models, the magnetic field and the turbulence are generated by collision of bubbles of the new phase on a length scale that is a sizeable fraction of the cosmological horizon at the moment of the phase transition. If the bubble walls propagate with near relativistic velocities, the kinetic energy density released as they sweep through the plasma can be comparable to the overall energy density of the Universe. This energy can be transferred to magnetic energy, and equipartition between magnetic energy and kinetic energy of the fluid bulk motion can be established by MHD turbulence. A SGWB would then be produced, sourced by the shear stresses arising from several processes: the MHD processed magnetic field, considered in this work, but also the associated kinetic turbulence, compressional velocity modes in the fluid \cite{Hindmarsh:2013xza,Hindmarsh:2015qta,hindmarsch17,Cutting:2019zws,Hindmarsh:2019phv}, and direct collision of the bubble walls \cite{Kosowsky:1991ua,Kosowsky:1992vn,Huber:2008hg,Cutting:2018tjt,Cutting:2020nla}.
The relative importance of these GW sources depends on the details of the phase transition. Bubble collisions are expected to dominate only for phase transitions with very strong supercooling. Sound waves are expected to dominate if the phase transition is weakly first order, and turbulence does not have time to develop within one Hubble time. MHD turbulence  likely dominates in the case of medium to strong phase transitions, when the shock timescale is shorter than the Hubble time. 
This represents the most natural scenario within which our result holds.

If interpreted as we propose in this work, the NANOGrav measurement would point to a first-order nature of the QCD phase transition, unless the magnetic field is generated at the QCD scale by some other, yet unknown, mechanism, not related to the presence of broken-phase bubbles. The order of the QCD phase transition might be altered in the Early Universe depending on the lepton asymmetry, as investigated in Refs. \cite{firstorderQCD,Wygas:2018otj}. PTA measurements of the SGWB originating from this phase transition can help to identify the beyond Standard Model physics effects converting a confinement cross-over into a first-order phase transition at the QCD temperature scale.

Furthermore, the interpretation of the NANOGrav measurement in terms of the SGWB from a magnetic field generated at the QCD phase transition has an implication in the context of the ``Hubble tension'' problem. Different measurements of the current expansion rate of the Universe, $H_0$, based on CMB probes \cite{planck_h0} and measurements in the local Universe \cite{shoes_h0,cosmograil_h0}, provide values which are inconsistent at $>4\sigma$ level (see, e.g., \cite{pogosyan} and references therein).
Reference \cite{pogosyan} has proposed that clumping of baryons in the primordial fluid induced by  magnetic field influences the CMB signal, relaxing the Hubble tension if the magnetic field parameters are within the range shown by the red interval in Fig. \ref{fig:exclusion}. In the figure, this interval is superimposed on the green shaded line representing the endpoint of the cosmological magnetic field evolution. 

Remarkably, the range of magnetic field amplitude and correlation scale suitable to release the Hubble tension according to Ref.~\cite{pogosyan} is exactly the one suitable to account for the NANOGrav detection if the magnetic field is generated at the QCD phase transition. The two field measurements are related by the dynamical evolutionary path of magnetic field amplitude and correlation length.
If the magnetic field generated at the QCD phase transition is non-helical, its strength and correlation length evolve following the $\tilde B\propto \tilde l_B^{-5/2}$ or $\tilde B\propto \tilde l_B^{-3/2}$ line determined by freely decaying turbulence \cite{banerjee,durrer13}. 
The  evolutionary path $\tilde B\propto \tilde l_B^{-3/2}$  shown by the rose arrow in Fig.~\ref{fig:exclusion} corresponds to compressible turbulence of the primordial plasma \cite{durrer13}. The range of possible endpoints of the cosmological evolution of the magnetic field, with parameters derived from the NANOGrav data, coincides with the range of magnetic field  parameters at the CMB epoch, derived in Ref. \cite{pogosyan}. 
This result is fairly independent on the uncertainties affecting the SGWB spectrum from MHD turbulence, that we have parameterized in terms of two free parameters. It holds if the SGWB characteristic frequency corresponds to the largest processed eddy scale, or to the energy-containing scale of the MHD turbulence, provided this latter is within one order of magnitude from the  horizon at the QCD time.

The model of magnetic field and GW production at the QCD phase transition analyzed here can be fully tested with the next generation PTA, supplemented with the SKA \cite{ska}. PTA will be able to provide high-confidence measurements of the SGWB possibly produced at the QCD epoch, at redshift $z\sim 10^{12}$. These can be combined with intergalactic magnetic field measurements by the next-generation gamma-ray telescope CTA \cite{cta,neronov09,Korochkin:2020pvg}, which has sufficient sensitivity to probe a relic magnetic field at $z=0$ (see Fig.~\ref{fig:exclusion}, \cite{Korochkin:2020pvg}). 
\vskip0.2cm
\section*{ACKNOWLEDGEMENTS}
This work is supported by the French National Research Agency (ANR) project MMUniverse (ANR-19-CE31-0020).

\bibliography{references}

\end{document}